\documentclass[twocolumn,secnumarabic,amssymb,nobibnotes, aps, prl]{revtex4}

\usepackage{graphicx}
\usepackage{natbib}
\usepackage{url}
\usepackage{textcase}
\usepackage{bm}
\usepackage{xcolor}
\usepackage{comment}
\usepackage{soul}
\def \ai{\textit{ab-initio}}
\def \bii{BiI$_{3}$}
\def \Fref{Fig.~\ref}
\begin{document}

\title{Photoinduced modulation of the excitonic resonance\\ via coupling with coherent phonons in a layered semiconductor}%

\author{S. Mor$^{1,2}$, V. Gosetti$^{1,2,9}$, A. Molina-S\'{a}nchez$^3$, D. Sangalli$^4$, S. Achilli$^5$, V. F. Agekyan$^7$ , P. Franceschini$^{1,2,8}$, C. Giannetti$^{1,2}$, L. Sangaletti$^{1,2}$,
S. Pagliara$^{1,2}$}

\email[corresponding author:]{stefania.pagliara@unicatt.it}

\affiliation{$^1$Department of Mathematics and Physics, Universit\`a Cattolica, I-25121 Brescia, Italy}
\affiliation{$^2$Interdisciplinary Laboratories for Advanced Materials Physics (I-LAMP), Universit\`a Cattolica, I-25121 Brescia, Italy}
 
\affiliation{$^3$Institute of Materials Science (ICMUV), University of Valencia,  Catedr\'{a}tico Beltr\'{a}n 2,  E-46980,  Valencia,  Spain}

\affiliation{$^4$Istituto di Struttura della Materia-CNR (ISM-CNR), Division of Ultrafast Processes in Materials (FLASHit), Area della Ricerca di Roma 1, Monterotondo Scalo, Italy}

\affiliation{$^5$ Dipartimento di Fisica, Universit\`a degli Studi
di Milano, via Celoria 6, 20133 Milano, Italy}

\affiliation{$^6$ CNR-ISTM and Dipartimento di Chimica, Universit\`a degli Studi di Milano}

\affiliation{$^7$ St. Petersburg State University, St. Petersburg, 199034, Russia}

\affiliation{$^8$Department of Physics and Astronomy, KU Leuven, Celestijnenlaan 200D, 3001 Leuven, Belgium}

\affiliation{$^9$ Department of Materials Engineering, KU Leuven, Kasteelpark Arenberg 44, 3001 Leuven, Belgium}

\date{\today}%

\begin{abstract}
The coupling of excitons with atomic vibrations plays a pivotal role on the nonequilibrium optical properties of layered semiconductors. However, addressing the dynamical interaction between excitons and phonons represents a hard task both experimentally and theoretically. By means of time-resolved broadband optical reflectivity combined with state-of-the-art ab-initio calculations of a bismuth triiodide single crystal, we unravel the universal spectral fingerprints of exciton--phonon coupling in layered semiconductors. Furthermore, we microscopically relate a photoinduced coherent energy modulation of the excitonic resonance to coherent optical phonons, thereby tracking the extent of the photoinduced atomic displacement in real-space. Our findings represent a step forward on the road to coherent manipulation of the excitonic properties on ultrafast timescales.
\end{abstract}

\maketitle
Electron--phonon coupling is among the fundamental interactions in condensed matter which govern the nonequilibrium optoelectronic properties of materials by, for instance, guiding the relaxation dynamics of quasiparticles \cite{Allen1987,Sangiovanni2006,Bauer2015,Sjakste2018} and assisting nonthermally-driven electronic phase transitions \cite{Hellmann2014,Porer2014,Maklar2021}. In semiconductors, the electron--hole Coulomb interaction also plays a crucial role in allowing the formation of bound states (excitons) upon photoabsorption. The binding energy of excitons is usually of a few meV, making them unstable at room temperature and difficult to be detected. However, in layered semiconductors, the reduced dielectric screening strongly enhances the exciton binding energy up to several hundreds of meV~\cite{Huser2013}. This fact, together with the technological relevance of layered materials, is triggering enormous research efforts in capturing exciton dynamics on the femto- to picoseconds timescale \cite{Carvalho2015,Paleari2019,Trovatello2020,Li2021,Brem2020,Chen2020}.
Recently, exciton--phonon coupling has been proposed to impact on not only the incoherent carrier-cooling dynamics, but also the homogeneous linewidth of the excitonic luminescence, and even the diffusion and the coherence length of excitons~\cite{Sanchez2017,Chen2020}.

While access to these complex dynamics can be a great challenge from experimentally and theoretically, their understanding and control is of paramount interest for the technological progress, as well as from a break-through knowledge perspective. Experimentally, resonant Raman scattering spectroscopy is among the most common techniques to address exciton--phonon coupling~\cite{Ganguly1967,Miranda2017}. However, the exciton dynamics remains out of reach due to the lack of temporal resolution. Time-resolved optical spectroscopy is the most suitable tool to investigate the interaction between excitons and phonons in the ultrafast time domain. By measuring transient changes in the optical polarization, the technique provides macroscopic snapshots of the nonequilibrium behavior of the system. On the theoretical side, \ai\, calculations of the exciton--phonon coupling combines density-functional-perturbation theory for the electron--phonon matrix elements~\cite{Giustino2017}, with the Bethe--Salpeter Equation (BSE), which correctly describes the excitonic physics~\cite{Onida1995}. Such scheme has been recently developed and applied to two-dimensional materials~\cite{Cudazzo2020,Chen2020} to compute exciton--phonon lifetimes. An alternative approach computes the excitonic properties solving the BSE within the GW approximation both at equilibrium and with the atoms displaced along specific phonon modes~\cite{Trovatello2020}. This method further enables an estimate of atomic displacements with exceptionally high spatial resolution on the sub-picometer scale~\cite{Katsuki2013}. 

In this Letter, by combining time-resolved broadband optical reflectivity measurements with state-of-the-art spinorial \ai--BSE calculations \cite{Marsili2021}, we unveil the spectral fingerprints of exciton--phonon coupling in a representative van-der-Waals-layered semiconductor, the bismuth tri-iodide (\bii) single crystal. Recently, \bii~has come into focus as an efficient non-toxic replacement of lead-based perovskites in optoelectronic applications \cite{Lehner2015, Shi2017}. The absorption spectrum is dominated by a direct excitonic resonance at 620~nm (2.00~eV) owing lifetime of nanoseconds and binding energy of hundreds of meV \cite{Kaifu1988}. Such large binding energy in a bulk semiconductor is a rather unique feature. Resonant Raman scattering combined with luminescence~\cite{Nila2017,SAITOH2000, KARASAWA1981} have identified an interaction between excitons and the main A$_g$ vibrational mode at 3.4~THz (113~cm$^{-1}$)~\cite{Tiwari2018}. More recently, transient absorption of spin-coated thin films of \bii\, suggested that coherent optical phonons and excitons should be coupled in the ultrafast time domain~\cite{Scholz2018}. All these aspects make \bii\, an ideal playground to explore the photophysics of exciton--phonon coupling in semiconductors. However, both the spectral fingerprint of the coupling and its impact on the exciton and phonon dynamics has remained elusive until now, likely due to the difficulty in providing univocal proof both experimentally and theoretically.

Here, we provide evidences that exciton--phonon coupling sets in at the photoexcitation time and triggers a coherent modulation of the excitonic resonance of \bii. The displacive photoexcitation of coherent optical phonons produces oscillations of the transient reflectivity which are enhanced by almost one order of magnitude at wavelengths close to the excitonic resonance. Moreover, the exciton energy varies periodically within a few picoseconds, in phase with the coherently-generated atomic vibrations. This is captured by retrieving the initial phase of the coherent phonon amplitude across the exciton wavelength and monitoring the energy shift of the excitonic peak. The \ai\, calculations corroborate that all observations are univocal and universal fingerprints of exciton--phonon coupling.  Moreover, our joint experimental and theoretical approach provides a unique tool to address exciton--phonon coupling in real space by retrieving the extent of the atomic displacement responsible for the coherent modulation of the excitonic resonance.

\begin{figure}
\includegraphics[width=1\columnwidth]{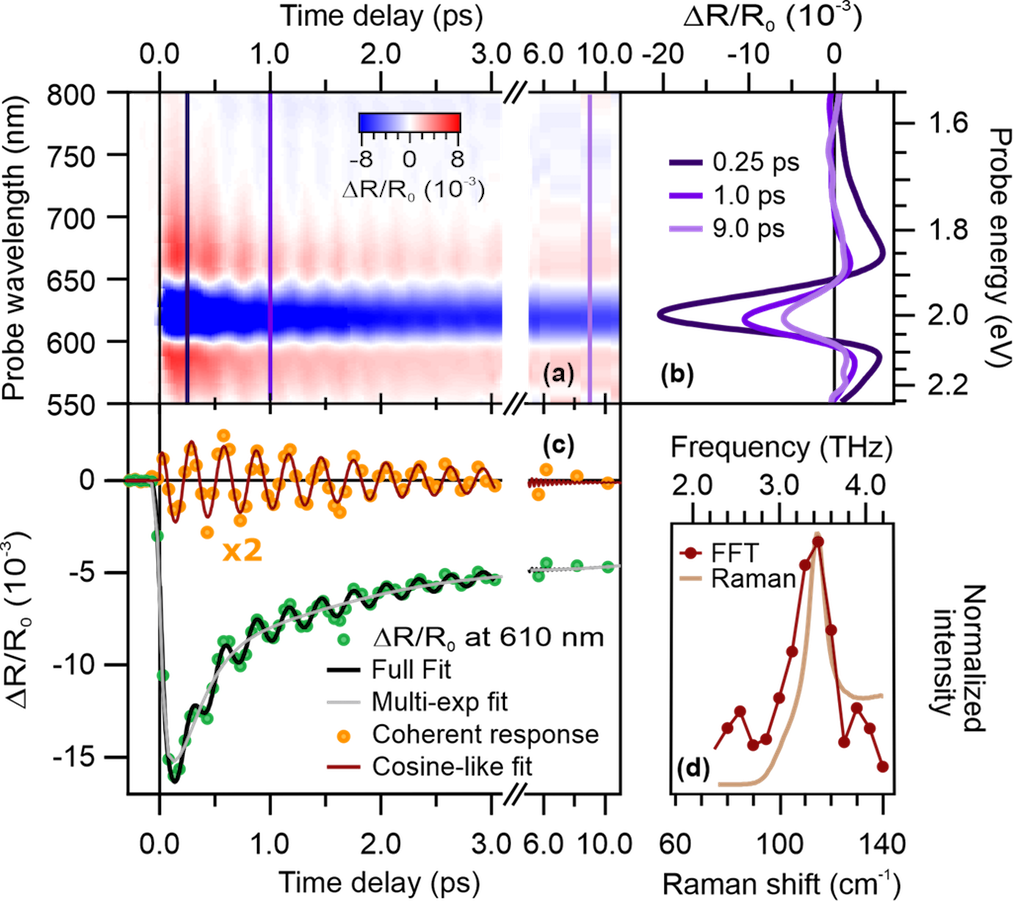}
\caption{(a) $\Delta R / R_0$ as a function of probe wavelength and pump-probe time delay. (b) Vertical linecuts of (a) at selected time delays of 0.25, 1.0 and 9.0~ps. (c) Horizontal linecut of (a) at 610~nm (green circles) superimposed by its full-fit curve (black line). The multi-exponential fit contribution (gray line) is subtracted from the data to obtain the bare coherent optical response (orange circles) shown together with the relevant fit  (red line). (d) FFT of the coherent optical response at 610~nm (red circles) and Raman-scattering spectrum (light brown line)}
\label{Fig1}
\end{figure}

The photoinduced transient reflectivity change, $(\Delta R / R_0) (t)$, is shown as color scale in \Fref{Fig1}(a) as a function of probe wavelength (left axis), probe photon energy (right axis), and pump-probe time delay (top axis). The measurement is carried out at room temperature and the pump photon energy is 2.78~eV (445~nm), well above the band gap of \bii. This enables the excitation of $n_c = 2 \times 10^{19}$ el/cm$^3$ quasi-free carriers into the conduction bands~\cite{Supplementary}. At positive time delays, i.e. after photoexcitation, a pronounced negative (blue) signal with minimum at 620~nm (2.00~eV) indicates a transient decrease of reflectivity ($(\Delta R / R_0) (t) < 0$) in the spectral range of the excitonic resonance. Above and below this spectral region, the signal is positive (red), as the transient reflectivity increases ($(\Delta R / R_0) (t) > 0$). The temporal evolution of both the negative and positive spectral features is emphasized by the spectra reported in \Fref{Fig1}(b) for selected time delays. At all probe wavelengths, this incoherent optical response is superimposed by a periodic intensity modulation (see, e.g. the coherent response at 610~nm in \Fref{Fig1}(c)) resulting from the generation of coherent optical phonons by the pump pulse. 

The transient reflectivity signal at each wavelength is fitted by a sum of four exponentially-decaying functions (incoherent optical response) and an exponentially-decaying cosine function (coherent optical response) convoluted with a Gaussian function (pump-pulse cross-correlation) \cite{Supplementary}. 
\Fref{Fig1}(c) shows the fit (black line) for the data at 610~nm (green circles). The gray line is the multi-exponential fit contribution to the incoherent optical response, only. At all wavelengths, we find that the latter decays through multiple relaxation channels with time constants on the order of 200~fs, few picoseconds, hundreds of picoseconds and 1.5~ns, respectively. These values agree with previous measurements of the transient absorption of spin-coated \bii\, thin films \cite{Scholz2018} and the luminescence of \bii\, single crystals \cite{Brandt2015}. Accordingly, we assign the fastest dynamics to carrier-carrier scattering, and the two intermediate dynamics to electron--phonon scattering processes and the beginning of electron-hole recombination.   
The slowest dynamics exceed the investigated temporal window and its time constant has been hold during the fitting procedure to the literature value~\cite{Brandt2015} of the electron-hole radiative recombination process.

After subtraction of the multi-exponential fit component (gray line in \Fref{Fig1}(c)), the coherent optical response (orange circles) is retrieved. Its fast Fourier transform (FFT, red circles  in \Fref{Fig1}(d)) reveals an oscillation at 3.41~$\pm$~0.02~THz matching the frequency of the A$_g$ phonon mode (light brown line)~\cite{Tiwari2018}. Moreover, the oscillation intensity is damped, on average, within 1.860~$\pm$~0.075~ps, which provides the dephasing time of the coherent phonons.

\begin{figure}
\includegraphics[width=1\columnwidth]{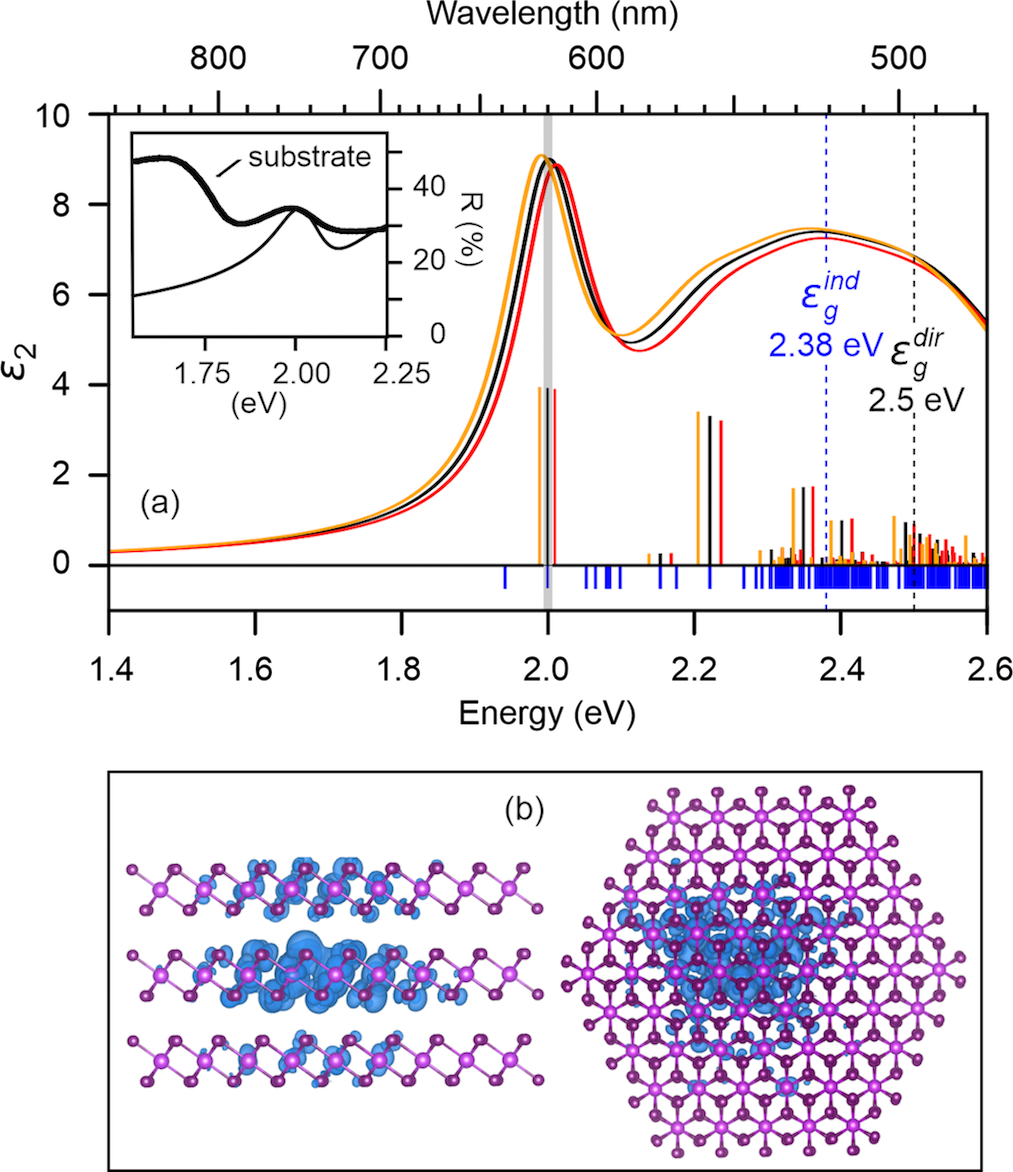}
\caption{(a) BSE absorption spectrum of \bii\, computed at equilibrium (black line) and with the atoms displaced along the $A_{g}$ phonon mode with $\pm\Delta_0$~=~0.05~Bohr $\approx$~2.6~pm (red and orange line). The corresponding equilibrium reflectivity is shown in the inset and compared with the experimental measurement. The blue vertical lines represent the excitonic energies at equilibrium, while the black, red, orange bars are the same poles weighted by the oscillator strengths. (b-c) Real-space representation of the exciton wavefunction.
}
\label{Fig2}
\end{figure}

To shed light on the nature of the transient optical response, we perform numerical analysis of the optical properties of \bii\, via \ai\,simulations within the GW+BSE scheme \cite{Supplementary}.
As shown in \Fref{Fig2}(a), the imaginary part of the dielectric function is dominated, in the optical region investigated in our measurements, by an excitonic peak with binding energy of 500~meV~\footnote{The exciton binding energy reported in the literature is 160-180~meV~\cite{Kaifu1988}. Such estimate is based on the Hydrogen model accounting for the macroscopic screening. Present calculations account for both macroscopic and microscopic screening effects.}. Such large binding energy is consistent with the strong in-plane localization of the exciton wavefunction~\cite{Habe2021} shown in the real-space representation of \Fref{Fig2}(b). The computed reflectivity spectrum, $R^{eq}(\omega)$ (thin line in inset of \Fref{Fig2}(a)) exhibits a resonance at the energy of the excitonic peak. To match $R^{eq}(\omega)$ with the experimental reflectance spectrum (thick line), the calculated peaks are shifted here by 0.35~eV. Thus, our calculations clearly show that the optical reflectivity is mostly dominated by the imaginary part of the dielectric function, in agreement with previous works~\cite{Jellison1999, Podraza2013}. We then interpret the incoherent optical response as follows:  the negative peak at 620~nm (2.00~eV) is assigned to the pump-induced photobleaching of the excitonic resonance; the positive signal at shorter and longer wavelengths to a combination of photoabsorption between transiently excited states,  screening-induced optical gap renormalization~\cite{Scholz2018}, and increase of the exciton linewidth.

\begin{figure}
\includegraphics[width=1\columnwidth]{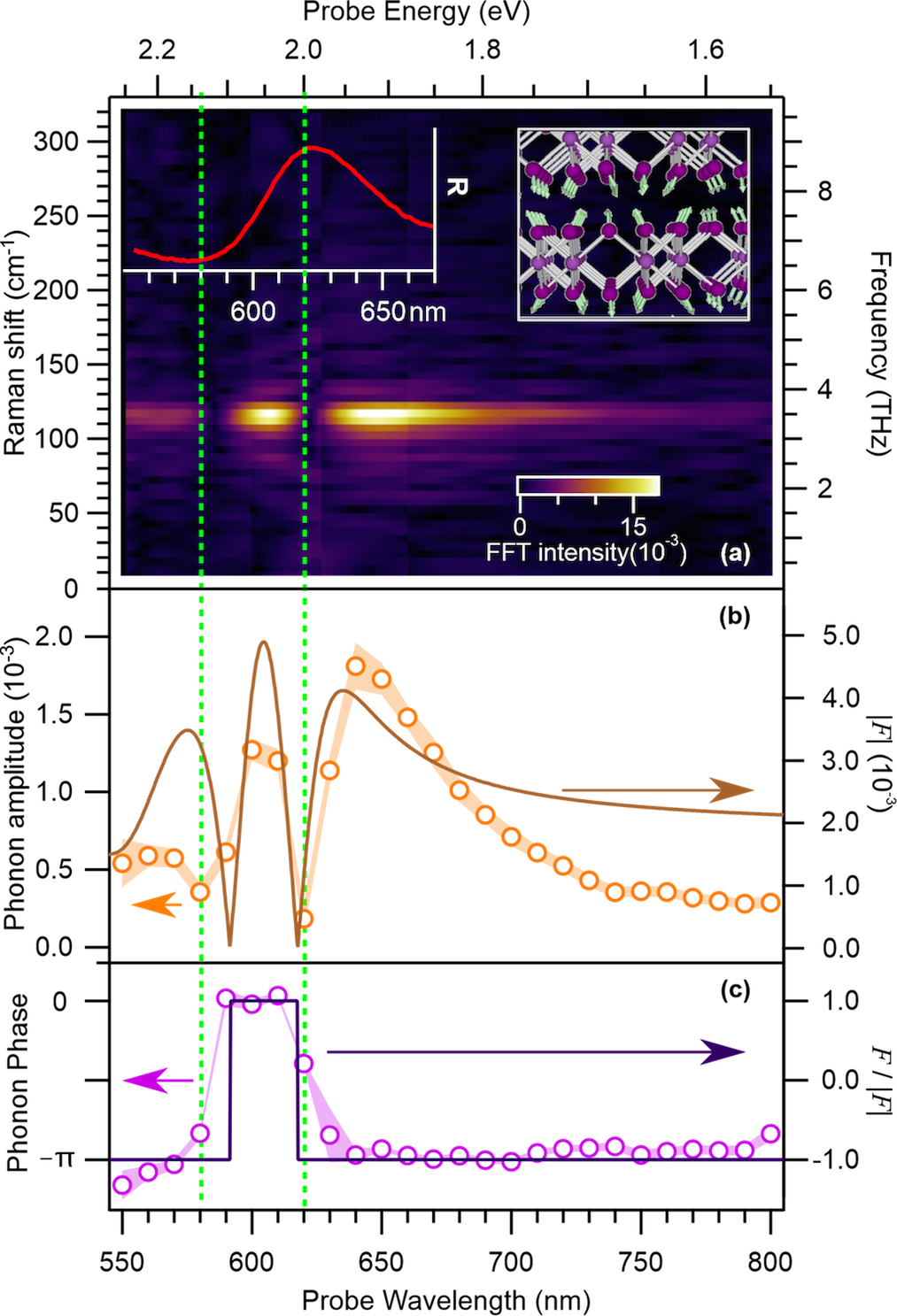}
\caption{(a) FFT of the coherent optical response at all probed wavelengths. Left inset: the measured equilibrium reflectance (as in inset of \Fref{Fig2}(a)). Right inset: the real-space representation of the A$_g$ phonon mode. (b) Experimental initial amplitude and (c) phase of the coherent optical phonon as function of the probe wavelength (left axis) and corresponding calculated quantities (right axis).}
\label{Fig3}
\end{figure}

We now focus on the bare coherent optical response of \bii. The frequency-resolved coherent optical phonon spectrum is obtained by FFT at each probe wavelength and plotted in color scale in \Fref{Fig3}(a). It exhibits one intense feature centered at 3.41~$\pm$~0.02~THz, i.e. at the frequency of the A$_g$ phonon mode whose calculated real-space representation is depicted in inset. A closer look at the FFT-spectrum intensity reveals two peculiarities: an enhancement in the exciton-wavelength region and a quenching at 580~nm (2.14~eV) and 620~nm (2.00~eV). We argue that both the enhancement and the quenching of the phonon intensity are unambiguous markers for exciton--phonon coupling in photoexcited semiconductors.

The fit parameters of the initial phonon amplitude and phase are plotted on the left axes of \Fref{Fig3}(b) and (c), respectively. The corresponding \ai\,quantities are explicitly calculated and reported on the right axis for direct comparison. As a first result, the phonon phase indicates, at all wavelengths, a cosine-like oscillation of the atoms around their equilibrium position. We deduce that coherent optical phonons are generated via a displacive excitation mechanism, consistent with the full symmetry of the A$_g$ vibrational mode and the absorptive excitation regime of our experiment~\cite{Cheng1991,Zeiger1992,Ishioka2006, Ishioka2010}.

The calculated amplitude and phase are obtained as follows. First, the GW+BSE reflectivity is re-computed with the atoms displaced by an amount ${\Delta X_{A_g}=\Delta_0\, u_{A_g}}$ along the direction of the eigenvector $u_{A_g}$ of the A$_g$ phonon mode (green arrows, inset of \Fref{Fig3}(a)). Assuming that the atoms oscillates according to ${X(t)=X_0+\Delta X_{A_g} cos(\omega_{A_g}t)}$, we obtain the reflectivity, $R^{X}(\omega)$, and thus the reflectivity variation, ${F(\omega)=(R^X-R^{eq})/R^{eq}}$, for distinct positions of the displaced atoms. We find that the atomic displacement induces opposite changes in the band gap and the exciton binding energy, respectively, resulting in a total shift of the excitonic peak, $\Delta E$. The overall picture is of an excitonic resonance that oscillates back and forth in energy as shown in \Fref{Fig2}(a) (yellow and red spectra).

We now compare the calculated 
$|F(\omega)|$ and $F(\omega)/|F(\omega)|$ 
with the measured initial amplitude and phase of the coherent optical response, respectively. Clearly, both  quantities confirm the amplitude enhancement as well as the two nodes at the wavelengths of the $\pi$-phase shift. The node at the exciton wavelength (620~nm) is due to opposite sign of the reflectivity variation generated by the shift of the excitonic peak. Concerning the other node, a bunch of excitons with strong exciton--phonon coupling is found to produce a broad peak centered at 525~nm (2.35~eV). Thus, the second node at 580~nm results from the combination of the displacements of the two excitonic peaks. All-in-all, we provide unambiguous proof of coupling between coherently activated A$_g$ phonons and excitons. Furthermore, by exploiting the excellent agreement between theory and experiment, we provide a quantitative estimate of the photoinduced atomic displacement and the associated energy shifts. An intensity variation of the coherent optical response on the order of $F(\omega)\approx 10^{-3}$ is obtained in the simulations upon atomic displacement in the range of $\Delta_0\approx 10^{-1}$~picometers. In turn, the excitonic peak is predicted to shift by $\Delta E\approx 10^{-1}$~meV.
\begin{figure}
\includegraphics[width=1\columnwidth]{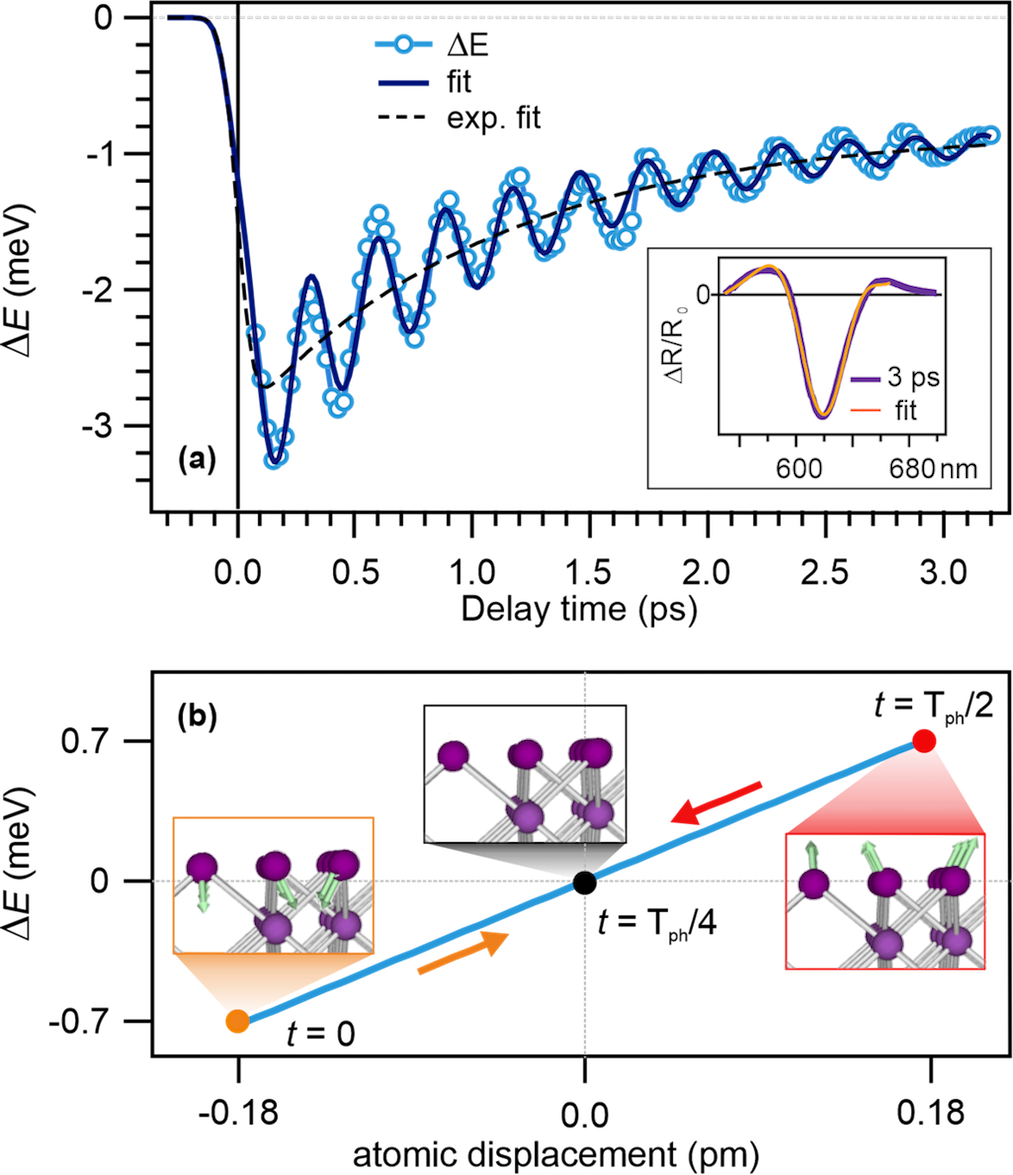}
\caption{(a) Energy shift of the excitonic peak (blue circles). The $\Delta E$ fit curve (solid line) is given by a single-exponential decay function (dashed line) multiplied by an exponentially-damped cosine function. In inset, spectrum at 3~ps (purple line) and relative differential fit (orange line). (b) The microscopic effect of exciton--phonon coupling: renormalization of the excitonic peak position by the atomic displacement.}
\label{Fig4}
\end{figure}

To unravel the impact of exciton--phonon coupling on the dynamics of the excitonic resonance, we follow the transient energy position of the excitonic peak. This is obtained by differential fitting of the transient reflectivity spectra (inset of \Fref{Fig4}(a)) after parametrization of the equilibrium reflectance with a set of Gaussian functions (see \cite{Supplementary} for details). \Fref{Fig4}(a) reports the energy shift, $\Delta E$, from the equilibrium position of the Gaussian peak accounting for the excitonic resonance (blue circles). The evaluated temporal window spans up to a few picoseconds to cover the phonon dephasing time. Upon photoexcitation, the excitonic resonance undergoes an instantaneous redshift (within the pump-pulse crosscorrelation) followed by a blueshift towards the equilibrium position. The latter evolves on a fast timescale of 1.110~$\pm$~0.075~ps (dashed line), and a slower one that exceeds the evaluated temporal window and is accounted in the fit by an additional offset. The blueshift is superimposed by a periodic energy modulation at the exact frequency of the A$_g$-phonon mode, damped on a timescale of 1.420~$\pm$~0.075~ps.  Both the red- and blueshifts  are rather common in photoexcited semiconductors and typically relate to  band-gap renormalization~\cite{Pagliara2011,Chernikov2015,Pogna2016,Mor2017} and exciton-binding-energy changes~\cite{Cunningham2017,Jiang2018} induced by dynamical screening, and conduction-band filling by photoexcited carriers \cite{Ullrich2014,Franceschini2020,Jnawali2020}. Conversely, the coherent modulation of the excitonic resonance has been rarely observed in the photoinduced dynamics of semiconductors with strong exciton--phonon coupling. We explain this fact by noting that a suitable intensity ratio of the coherent and incoherent optical response is required in order to avoid the exciton energy modulation to be buried underneath the photobleaching spectral intensity, as reported, for instance, in \cite{Trovatello2020}. Our measurements indicate that such ratio is on the order of 1:10 and is achieved by off-resonance photoexcitation with energy above the optical band gap in order to retain from too intense photobleaching of the excitonic resonance. 

The periodic modulation of the excitonic peak exhibits an initial amplitude of $\approx$~0.7~meV~\footnote{We note that a periodic modulation of the transient reflectivity minimum is already visible by eye. The high statistics of our transient reflectivity data allows us to resolve relative energy shifts of only few tens of meV which are robust against the choice of fit analysis.} reproduced by the simulation with an atomic displacement of 0.18~pm (see \Fref{Fig4}(b)). The corresponding reflectivity variation is on the expected order of $10^{-3}$, differing only by a factor of $\approx 2$ form the measured phonon amplitude at the respective probe wavelengths (see \Fref{Fig3}(b)). All these findings demonstrate that tracking the energy modulation of the excitonic resonance induced by coherent phonons enables to trace atomic displacements along the direction of the phonon mode. Effectively, our combined experimental and theoretical approach successfully addresses the coherent phonon dynamics coupled to the exciton in real-space with sub-picometer resolution.

In conclusion, the amplitude enhancement and phase shift of the coherent phonon are established as univocal spectral fingerprints of exciton--phonon coupling. We prove that the excitonic resonance of the layered semiconductor \bii\,can be optically modulated via coupling of coherent atomic vibrations of which we are able to estimate the spatial extent with sub-picometer resolution. Noteworthly, the present results can be extended to any photoexcited semiconductor with the exciton strongly coupled to a lattice deformation. As a perspective, we propose that ultrafast modulation of excitonic resonances could be experimentally controlled by varying the laser fluence or tuning the pump photon energy across the exciton transition, paving the way towards optical control of the nonequilibrium optoelectronic properties of low-dimensional semiconductors.

\section*{Acknowledgements}
S.P., L.S. and C.G. acknowledge partial support from D.1 and D.2.2 grants of the Universit\'a Cattolica del Sacro Cuore. C.G. and P.F. acknowledge financial support from MIUR through the project CENTRAL  (Prot. 20172H2SC4\_005) within the PRIN 2017 program. D.S. acknowledges the funding received from MIUR PRIN BIOX Grant No. 20173B72NB, and from the European Union projects: MaX Materials design at the eXascale H2020-INFRAEDI-2018-2020, Grant agreement No.824143; Nanoscience Foundries and Fine Analysis - Europe H2020-INFRAIA-2014-2015, Grant agreement No. 654360.


\section*{References}
\begin{thebibliography}{48}
\expandafter\ifx\csname natexlab\endcsname\relax\def\natexlab#1{#1}\fi
\expandafter\ifx\csname bibnamefont\endcsname\relax
  \def\bibnamefont#1{#1}\fi
\expandafter\ifx\csname bibfnamefont\endcsname\relax
  \def\bibfnamefont#1{#1}\fi
\expandafter\ifx\csname citenamefont\endcsname\relax
  \def\citenamefont#1{#1}\fi
\expandafter\ifx\csname url\endcsname\relax
  \def\url#1{\texttt{#1}}\fi
\expandafter\ifx\csname urlprefix\endcsname\relax\def\urlprefix{URL }\fi
\providecommand{\bibinfo}[2]{#2}
\providecommand{\eprint}[2][]{\url{#2}}

\bibitem[{\citenamefont{Allen}(1987)}]{Allen1987}
\bibinfo{author}{\bibfnamefont{P.~B.} \bibnamefont{Allen}},
  \bibinfo{journal}{Phys. Rev. Lett.} \textbf{\bibinfo{volume}{59}},
  \bibinfo{pages}{1460} (\bibinfo{year}{1987}).
 % \urlprefix\url{https://link.aps.org/doi/10.1103/PhysRevLett.59.1460}.

\bibitem[{\citenamefont{Sangiovanni et~al.}(2006)\citenamefont{Sangiovanni,
  Capone, and Castellani}}]{Sangiovanni2006}
\bibinfo{author}{\bibfnamefont{G.}~\bibnamefont{Sangiovanni}},
  \bibinfo{author}{\bibfnamefont{M.}~\bibnamefont{Capone}}, \bibnamefont{and}
  \bibinfo{author}{\bibfnamefont{C.}~\bibnamefont{Castellani}},
  \bibinfo{journal}{Phys. Rev. B} \textbf{\bibinfo{volume}{73}},
  \bibinfo{pages}{165123} (\bibinfo{year}{2006}).
 % \urlprefix\url{https://link.aps.org/doi/10.1103/PhysRevB.73.165123}.

\bibitem[{\citenamefont{Bauer et~al.}(2015)\citenamefont{Bauer, Marienfeld, and
  Aeschlimann}}]{Bauer2015}
    \bibinfo{author}{\bibfnamefont{M.}~\bibnamefont{Bauer}},
  \bibinfo{author}{\bibfnamefont{A.}~\bibnamefont{Marienfeld}},
  \bibnamefont{and}
  \bibinfo{author}{\bibfnamefont{M.}~\bibnamefont{Aeschlimann}},
  \bibinfo{journal}{Progress in Surface Science} \textbf{\bibinfo{volume}{90}},
  \bibinfo{pages}{319 } (\bibinfo{year}{2015}), ISSN \bibinfo{issn}{0079-6816}.
  %\urlprefix\url{http://www.sciencedirect.com/science/article/pii/S0079681615000192}.

\bibitem[{\citenamefont{Sjakste et~al.}(2018)\citenamefont{Sjakste, Tanimura,
  Barbarino, Perfetti, and Vast}}]{Sjakste2018}
\bibinfo{author}{\bibfnamefont{J.}~\bibnamefont{Sjakste}},
  \bibinfo{author}{\bibfnamefont{K.}~\bibnamefont{Tanimura}},
  \bibinfo{author}{\bibfnamefont{G.}~\bibnamefont{Barbarino}},
  \bibinfo{author}{\bibfnamefont{L.}~\bibnamefont{Perfetti}}, \bibnamefont{and}
  \bibinfo{author}{\bibfnamefont{N.}~\bibnamefont{Vast}},
  \bibinfo{journal}{Journal of Physics: Condensed Matter}
  \textbf{\bibinfo{volume}{30}}, \bibinfo{pages}{353001}
  (\bibinfo{year}{2018}).
 % \urlprefix\url{https://doi.org/10.1088/1361-648x/aad487}.

\bibitem[{\citenamefont{Hellmann et~al.}(2012)\citenamefont{Hellmann, Rohwer,
  Kall\"{a}ne, Hanff, Sohrt, Stange, Carr, Murnane, Kapteyn, Kipp
  et~al.}}]{Hellmann2014}
\bibinfo{author}{\bibfnamefont{S.}~\bibnamefont{Hellmann}},
  \bibinfo{author}{\bibfnamefont{T.}~\bibnamefont{Rohwer}},
  \bibinfo{author}{\bibfnamefont{M.}~\bibnamefont{Kall\"{a}ne}},
  \bibinfo{author}{\bibfnamefont{K.}~\bibnamefont{Hanff}},
  \bibinfo{author}{\bibfnamefont{C.}~\bibnamefont{Sohrt}},
  \bibinfo{author}{\bibfnamefont{A.}~\bibnamefont{Stange}},
  \bibinfo{author}{\bibfnamefont{A.}~\bibnamefont{Carr}},
  \bibinfo{author}{\bibfnamefont{M.}~\bibnamefont{Murnane}},
  \bibinfo{author}{\bibfnamefont{H.}~\bibnamefont{Kapteyn}},
  \bibinfo{author}{\bibfnamefont{L.}~\bibnamefont{Kipp}}, \bibnamefont{et~al.},
  \bibinfo{journal}{Nat. Comm.} \textbf{\bibinfo{volume}{3}}
  (\bibinfo{year}{2012}).
  %\urlprefix\url{https://doi.org/10.1038/ncomms2078}.

\bibitem[{\citenamefont{Porer et~al.}(2014)\citenamefont{Porer, Leierseder,
  Ménard, Dachraoui, Mouchliadis, Perakis, Heinzmann, Demsar, Rossnagel, and
  Huber}}]{Porer2014}
\bibinfo{author}{\bibfnamefont{M.}~\bibnamefont{Porer}},
  \bibinfo{author}{\bibfnamefont{U.}~\bibnamefont{Leierseder}},
  \bibinfo{author}{\bibfnamefont{J.-M.} \bibnamefont{Ménard}},
  \bibinfo{author}{\bibfnamefont{H.}~\bibnamefont{Dachraoui}},
  \bibinfo{author}{\bibfnamefont{L.}~\bibnamefont{Mouchliadis}},
  \bibinfo{author}{\bibfnamefont{I.~E.} \bibnamefont{Perakis}},
  \bibinfo{author}{\bibfnamefont{U.}~\bibnamefont{Heinzmann}},
  \bibinfo{author}{\bibfnamefont{J.}~\bibnamefont{Demsar}},
  \bibinfo{author}{\bibfnamefont{K.}~\bibnamefont{Rossnagel}},
  \bibnamefont{and} \bibinfo{author}{\bibfnamefont{R.}~\bibnamefont{Huber}},
  \bibinfo{journal}{Nat. Mat.} \textbf{\bibinfo{volume}{13}}
  (\bibinfo{year}{2014}).
  %\urlprefix\url{https://doi.org/10.1038/nmat4042}.

\bibitem[{\citenamefont{Maklar et~al.}(2021)\citenamefont{Maklar, Windsor,
  Nicholson, Puppin, Walmsley, Esposito, Porer, Rittmann, Leuenberger, Kubli
  et~al.}}]{Maklar2021}
\bibinfo{author}{\bibfnamefont{J.}~\bibnamefont{Maklar}},
  \bibinfo{author}{\bibfnamefont{Y.~W.} \bibnamefont{Windsor}},
  \bibinfo{author}{\bibfnamefont{C.~W.} \bibnamefont{Nicholson}},
  \bibinfo{author}{\bibfnamefont{M.}~\bibnamefont{Puppin}},
  \bibinfo{author}{\bibfnamefont{P.}~\bibnamefont{Walmsley}},
  \bibinfo{author}{\bibfnamefont{V.}~\bibnamefont{Esposito}},
  \bibinfo{author}{\bibfnamefont{M.}~\bibnamefont{Porer}},
  \bibinfo{author}{\bibfnamefont{J.}~\bibnamefont{Rittmann}},
  \bibinfo{author}{\bibfnamefont{D.}~\bibnamefont{Leuenberger}},
  \bibinfo{author}{\bibfnamefont{M.}~\bibnamefont{Kubli}},
  \bibnamefont{et~al.}, \bibinfo{journal}{Nat. Comm.}
  \textbf{\bibinfo{volume}{12}} (\bibinfo{year}{2021})-.
%  \urlprefix\url{https://doi.org/10.1038/s41467-021-22778-w}.

\bibitem[{\citenamefont{H\"user et~al.}(2013)\citenamefont{H\"user, Olsen, and
  Thygesen}}]{Huser2013}
\bibinfo{author}{\bibfnamefont{F.}~\bibnamefont{H\"user}},
  \bibinfo{author}{\bibfnamefont{T.}~\bibnamefont{Olsen}}, \bibnamefont{and}
  \bibinfo{author}{\bibfnamefont{K.~S.} \bibnamefont{Thygesen}},
  \bibinfo{journal}{Phys. Rev. B} \textbf{\bibinfo{volume}{88}},
  \bibinfo{pages}{245309} (\bibinfo{year}{2013}).
  %\urlprefix\url{https://link.aps.org/doi/10.1103/PhysRevB.88.245309}.

\bibitem[{\citenamefont{Carvalho et~al.}(2015)\citenamefont{Carvalho, Malard,
  Alves, Fantini, and Pimenta}}]{Carvalho2015}
\bibinfo{author}{\bibfnamefont{B.~R.} \bibnamefont{Carvalho}},
  \bibinfo{author}{\bibfnamefont{L.~M.} \bibnamefont{Malard}},
  \bibinfo{author}{\bibfnamefont{J.~M.} \bibnamefont{Alves}},
  \bibinfo{author}{\bibfnamefont{C.}~\bibnamefont{Fantini}}, \bibnamefont{and}
  \bibinfo{author}{\bibfnamefont{M.~A.} \bibnamefont{Pimenta}},
  \bibinfo{journal}{Phys. Rev. Lett.} \textbf{\bibinfo{volume}{114}},
  \bibinfo{pages}{136403} (\bibinfo{year}{2015}).
  %\urlprefix\url{https://link.aps.org/doi/10.1103/PhysRevLett.114.136403}.

\bibitem[{\citenamefont{Paleari et~al.}(2019)\citenamefont{Paleari, Miranda,
Molina-Sanchez, Wirtz}}]{Paleari2019} 
  \bibinfo{author}{\bibfnamefont{F.}~\bibnamefont{Paleari}},
  \bibinfo{author}{\bibfnamefont{H.}~\bibnamefont{P.~C.~Miranda}},
  \bibinfo{author}{\bibfnamefont{A.}~\bibnamefont{Molina-S\'anchez}},
  \bibnamefont{and} \bibinfo{author}{\bibfnamefont{L.}~\bibnamefont{Wirtz}},
  \bibinfo{journal}{Phys. Rev. Lett.} \textbf{\bibinfo{volume}{122}},
  \bibinfo{pages}{187401} (\bibinfo{year}{2019}).
%  \urlprefix\url{https://link.aps.org/doi/10.1103/PhysRevLett.122.187401}.

\bibitem[{\citenamefont{Trovatello et~al.}(2020)\citenamefont{Trovatello,
  Miranda, Molina-Sánchez, Borrego-Varillas, Manzoni, Moretti, Ganzer, Maiuri,
  Wang, Dumcenco et~al.}}]{Trovatello2020}
\bibinfo{author}{\bibfnamefont{C.}~\bibnamefont{Trovatello}},
  \bibinfo{author}{\bibfnamefont{H.~P.~C.} \bibnamefont{Miranda}},
  \bibinfo{author}{\bibfnamefont{A.}~\bibnamefont{Molina-Sánchez}},
  \bibinfo{author}{\bibfnamefont{R.}~\bibnamefont{Borrego-Varillas}},
  \bibinfo{author}{\bibfnamefont{C.}~\bibnamefont{Manzoni}},
  \bibinfo{author}{\bibfnamefont{L.}~\bibnamefont{Moretti}},
  \bibinfo{author}{\bibfnamefont{L.}~\bibnamefont{Ganzer}},
  \bibinfo{author}{\bibfnamefont{M.}~\bibnamefont{Maiuri}},
  \bibinfo{author}{\bibfnamefont{J.}~\bibnamefont{Wang}},
  \bibinfo{author}{\bibfnamefont{D.}~\bibnamefont{Dumcenco}},
  \bibnamefont{et~al.}, \bibinfo{journal}{ACS Nano}
  \textbf{\bibinfo{volume}{14}}, \bibinfo{pages}{5700} (\bibinfo{year}{2020}).
%  \urlprefix\url{https://doi.org/10.1021/acsnano.0c00309}.

\bibitem[{\citenamefont{Li et~al.}(2021)\citenamefont{Li, Trovatello,
  Dal~Conte, Nu{\ss}, Soavi, Wang, Ferrari, Cerullo, and Brixner}}]{Li2021}
\bibinfo{author}{\bibfnamefont{D.}~\bibnamefont{Li}},
  \bibinfo{author}{\bibfnamefont{C.}~\bibnamefont{Trovatello}},
  \bibinfo{author}{\bibfnamefont{S.}~\bibnamefont{Dal~Conte}},
  \bibinfo{author}{\bibfnamefont{M.}~\bibnamefont{Nu{\ss}}},
  \bibinfo{author}{\bibfnamefont{G.}~\bibnamefont{Soavi}},
  \bibinfo{author}{\bibfnamefont{G.}~\bibnamefont{Wang}},
  \bibinfo{author}{\bibfnamefont{A.~C.} \bibnamefont{Ferrari}},
  \bibinfo{author}{\bibfnamefont{G.}~\bibnamefont{Cerullo}}, \bibnamefont{and}
  \bibinfo{author}{\bibfnamefont{T.}~\bibnamefont{Brixner}},
  \bibinfo{journal}{Nat. Comm.} \textbf{\bibinfo{volume}{12}},
  \bibinfo{pages}{954} (\bibinfo{year}{2021}), ISSN \bibinfo{issn}{2041-1723}.
 % \urlprefix\url{https://doi.org/10.1038/s41467-021-20895-0}.

\bibitem[{\citenamefont{Brem et~al.}(2020)\citenamefont{Brem, Ekman,
  Christiansen, Katsch, Selig, Robert, Marie, Urbaszek, Knorr, and
  Malic}}]{Brem2020}
\bibinfo{author}{\bibfnamefont{S.}~\bibnamefont{Brem}},
  \bibinfo{author}{\bibfnamefont{A.}~\bibnamefont{Ekman}},
  \bibinfo{author}{\bibfnamefont{D.}~\bibnamefont{Christiansen}},
  \bibinfo{author}{\bibfnamefont{F.}~\bibnamefont{Katsch}},
  \bibinfo{author}{\bibfnamefont{M.}~\bibnamefont{Selig}},
  \bibinfo{author}{\bibfnamefont{C.}~\bibnamefont{Robert}},
  \bibinfo{author}{\bibfnamefont{X.}~\bibnamefont{Marie}},
  \bibinfo{author}{\bibfnamefont{B.}~\bibnamefont{Urbaszek}},
  \bibinfo{author}{\bibfnamefont{A.}~\bibnamefont{Knorr}}, \bibnamefont{and}
  \bibinfo{author}{\bibfnamefont{E.}~\bibnamefont{Malic}},
  \bibinfo{journal}{Nano Lett.} \textbf{\bibinfo{volume}{20}},
  \bibinfo{pages}{2849} (\bibinfo{year}{2020}).
%  \urlprefix\url{https://doi.org/10.1021/acs.nanolett.0c00633}.
  
\bibitem[{\citenamefont{Chen et~al.}(2020)\citenamefont{Chen, Sangalli, and
  Bernardi}}]{Chen2020}
\bibinfo{author}{\bibfnamefont{H.-Y.} \bibnamefont{Chen}},
  \bibinfo{author}{\bibfnamefont{D.}~\bibnamefont{Sangalli}}, \bibnamefont{and}
  \bibinfo{author}{\bibfnamefont{M.}~\bibnamefont{Bernardi}},
  \bibinfo{journal}{Phys. Rev. Lett.} \textbf{\bibinfo{volume}{125}},
  \bibinfo{pages}{107401} (\bibinfo{year}{2020}).
%  \urlprefix\url{https://doi.org/10.1103/PhysRevLett.125.107401}.

\bibitem[{\citenamefont{Molina-Sánchez
  et~al.}(2017)\citenamefont{Molina-Sánchez, Sangalli, Wirtz, and
  Marini}}]{Sanchez2017}
\bibinfo{author}{\bibfnamefont{A.}~\bibnamefont{Molina-Sánchez}},
  \bibinfo{author}{\bibfnamefont{D.}~\bibnamefont{Sangalli}},
  \bibinfo{author}{\bibfnamefont{L.}~\bibnamefont{Wirtz}}, \bibnamefont{and}
  \bibinfo{author}{\bibfnamefont{A.}~\bibnamefont{Marini}},
  \bibinfo{journal}{Nano Letters} \textbf{\bibinfo{volume}{17}},
  \bibinfo{pages}{4549} (\bibinfo{year}{2017}).
%  \urlprefix\url{https://doi.org/10.1021/acs.nanolett.7b00175}.

\bibitem[{\citenamefont{Ganguly and Birman}(1967)}]{Ganguly1967}
\bibinfo{author}{\bibfnamefont{A.~K.} \bibnamefont{Ganguly}} \bibnamefont{and}
  \bibinfo{author}{\bibfnamefont{J.~L.} \bibnamefont{Birman}},
  \bibinfo{journal}{Phys. Rev.} \textbf{\bibinfo{volume}{162}},
  \bibinfo{pages}{806} (\bibinfo{year}{1967}).
%  \urlprefix\url{https://doi.org/10.1103/PhysRev.162.806}.

\bibitem[{\citenamefont{Miranda et~al.}(2017)\citenamefont{Miranda, Reichardt,
  Froehlicher, Molina-Sánchez, Berciaud, and Wirtz}}]{Miranda2017}
  \bibinfo{author}{\bibfnamefont{H.~P.~C.} \bibnamefont{Miranda}},
  \bibinfo{author}{\bibfnamefont{S.}~\bibnamefont{Reichardt}},
  \bibinfo{author}{\bibfnamefont{G.}~\bibnamefont{Froehlicher}},
  \bibinfo{author}{\bibfnamefont{A.}~\bibnamefont{Molina-Sánchez}},
  \bibinfo{author}{\bibfnamefont{S.}~\bibnamefont{Berciaud}}, \bibnamefont{and}
  \bibinfo{author}{\bibfnamefont{L.}~\bibnamefont{Wirtz}},
  \bibinfo{journal}{Nano Lett.} \textbf{\bibinfo{volume}{17}}
  (\bibinfo{year}{2017}), ISSN \bibinfo{issn}{1530-6984}.
%  \urlprefix\url{https://doi.org/10.1021/acs.nanolett.6b05345}.

\bibitem[{\citenamefont{Giustino}(2017)}]{Giustino2017}
  \bibinfo{author}{\bibfnamefont{F.}~\bibnamefont{Giustino}},
  \bibinfo{journal}{Rev. Mod. Phys.} \textbf{\bibinfo{volume}{89}},
  \bibinfo{pages}{015003} (\bibinfo{year}{2017}).
%  \urlprefix\url{https://link.aps.org/doi/10.1103/RevModPhys.89.015003}.

\bibitem[{\citenamefont{Onida et~al.}(1995)\citenamefont{Onida, Reining, Godby,
  Del~Sole, and Andreoni}}]{Onida1995}
\bibinfo{author}{\bibfnamefont{G.}~\bibnamefont{Onida}},
  \bibinfo{author}{\bibfnamefont{L.}~\bibnamefont{Reining}},
  \bibinfo{author}{\bibfnamefont{R.~W.} \bibnamefont{Godby}},
  \bibinfo{author}{\bibfnamefont{R.}~\bibnamefont{Del~Sole}}, \bibnamefont{and}
  \bibinfo{author}{\bibfnamefont{W.}~\bibnamefont{Andreoni}},
  \bibinfo{journal}{Phys. Rev. Lett.} \textbf{\bibinfo{volume}{75}},
  \bibinfo{pages}{818} (\bibinfo{year}{1995}).
%  \urlprefix\url{https://link.aps.org/doi/10.1103/PhysRevLett.75.818}.

\bibitem[{\citenamefont{Cudazzo}(2020)}]{Cudazzo2020}
\bibinfo{author}{\bibfnamefont{P.}~\bibnamefont{Cudazzo}},
  \bibinfo{journal}{Phys. Rev. B} \textbf{\bibinfo{volume}{102}},
  \bibinfo{pages}{045136} (\bibinfo{year}{2020}).
%  \urlprefix\url{https://link.aps.org/doi/10.1103/PhysRevB.102.045136}.

\bibitem[{\citenamefont{Katsuki et~al.}(2013)\citenamefont{Katsuki, Delagnes,
  Hosaka, Ishioka, Chiba, Zijlstra, Garcia, Takahashi, Watanabe, Kitajima
  et~al.}}]{Katsuki2013}
\bibinfo{author}{\bibfnamefont{H.}~\bibnamefont{Katsuki}},
  \bibinfo{author}{\bibfnamefont{J.~C.} \bibnamefont{Delagnes}},
  \bibinfo{author}{\bibfnamefont{K.}~\bibnamefont{Hosaka}},
  \bibinfo{author}{\bibfnamefont{K.}~\bibnamefont{Ishioka}},
  \bibinfo{author}{\bibfnamefont{H.}~\bibnamefont{Chiba}},
  \bibinfo{author}{\bibfnamefont{E.~S.} \bibnamefont{Zijlstra}},
  \bibinfo{author}{\bibfnamefont{M.~E.} \bibnamefont{Garcia}},
  \bibinfo{author}{\bibfnamefont{H.}~\bibnamefont{Takahashi}},
  \bibinfo{author}{\bibfnamefont{K.}~\bibnamefont{Watanabe}},
  \bibinfo{author}{\bibfnamefont{M.}~\bibnamefont{Kitajima}},
  \bibnamefont{et~al.}, \bibinfo{journal}{Nat. Comm.}
  \textbf{\bibinfo{volume}{4}}, \bibinfo{pages}{2801} (\bibinfo{year}{2013}).
%  \urlprefix\url{https://doi.org/10.1038/ncomms3801}.

\bibitem[{\citenamefont{Marsili et~al.}(2021)\citenamefont{Marsili,
  Molina-S\'anchez, Palummo, Sangalli, and Marini}}]{Marsili2021}
\bibinfo{author}{\bibfnamefont{M.}~\bibnamefont{Marsili}},
  \bibinfo{author}{\bibfnamefont{A.}~\bibnamefont{Molina-S\'anchez}},
  \bibinfo{author}{\bibfnamefont{M.}~\bibnamefont{Palummo}},
  \bibinfo{author}{\bibfnamefont{D.}~\bibnamefont{Sangalli}}, \bibnamefont{and}
  \bibinfo{author}{\bibfnamefont{A.}~\bibnamefont{Marini}},
  \bibinfo{journal}{Phys. Rev. B} \textbf{\bibinfo{volume}{103}},
  \bibinfo{pages}{155152} (\bibinfo{year}{2021}).
%  \urlprefix\url{https://link.aps.org/doi/10.1103/PhysRevB.103.155152}.

\bibitem[{\citenamefont{Lehner et~al.}(2015)\citenamefont{Lehner, Wang, Fabini,
  Liman, Hébert, Perry, Wang, Chabinyc, and Seshadri}}]{Lehner2015}
\bibinfo{author}{\bibfnamefont{A.~J.} \bibnamefont{Lehner}},
  \bibinfo{author}{\bibfnamefont{H.}~\bibnamefont{Wang}},
  \bibinfo{author}{\bibfnamefont{D.~H.} \bibnamefont{Fabini}},
  \bibinfo{author}{\bibfnamefont{C.}~\bibnamefont{Liman}},
  \bibinfo{author}{\bibfnamefont{C.}~\bibnamefont{Hébert}},
  \bibinfo{author}{\bibfnamefont{E.}~\bibnamefont{Perry}},
  \bibinfo{author}{\bibfnamefont{G.}~\bibnamefont{Wang},
  \bibfnamefont{M.~Bazan}},
  \bibinfo{author}{\bibfnamefont{M.}~\bibnamefont{Chabinyc}}, \bibnamefont{and}
  \bibinfo{author}{\bibfnamefont{R.}~\bibnamefont{Seshadri}},
  \bibinfo{journal}{Appl. Phys. Lett.} \textbf{\bibinfo{volume}{107}},
  \bibinfo{pages}{131109} (\bibinfo{year}{2015}).
%  \urlprefix\url{https://doi.org/10.1063/1.4932129}.

\bibitem[{\citenamefont{Shi et~al.}(2017)\citenamefont{Shi, Guo, Chen, Li, Pan,
  Zhang, Xia, and Huang}}]{Shi2017}
\bibinfo{author}{\bibfnamefont{Z.}~\bibnamefont{Shi}},
  \bibinfo{author}{\bibfnamefont{J.}~\bibnamefont{Guo}},
  \bibinfo{author}{\bibfnamefont{Y.}~\bibnamefont{Chen}},
  \bibinfo{author}{\bibfnamefont{Q.}~\bibnamefont{Li}},
  \bibinfo{author}{\bibfnamefont{Y.}~\bibnamefont{Pan}},
  \bibinfo{author}{\bibfnamefont{H.}~\bibnamefont{Zhang}},
  \bibinfo{author}{\bibfnamefont{Y.}~\bibnamefont{Xia}}, \bibnamefont{and}
  \bibinfo{author}{\bibfnamefont{W.}~\bibnamefont{Huang}},
  \bibinfo{journal}{Adv. Mater.} \textbf{\bibinfo{volume}{29}},
  \bibinfo{pages}{1605005} (\bibinfo{year}{2017}).
%  \urlprefix\url{https://doi.org/10.1002/adma.201605005}.

\bibitem[{\citenamefont{Kaifu}(1988)}]{Kaifu1988}
\bibinfo{author}{\bibfnamefont{Y.}~\bibnamefont{Kaifu}},
  \bibinfo{journal}{Journal of Luminescence} \textbf{\bibinfo{volume}{42}},
  \bibinfo{pages}{61 } (\bibinfo{year}{1988}).
%  \urlprefix\url{https://doi.org/10.1016/0022-2313(88)90045-2}.

\bibitem[{\citenamefont{Nila et~al.}(2017)\citenamefont{Nila, Matea, Baibarac,
  and Baltog}}]{Nila2017}
\bibinfo{author}{\bibfnamefont{A.}~\bibnamefont{Nila}},
  \bibinfo{author}{\bibfnamefont{A.}~\bibnamefont{Matea}},
  \bibinfo{author}{\bibfnamefont{M.}~\bibnamefont{Baibarac}}, \bibnamefont{and}
  \bibinfo{author}{\bibfnamefont{I.}~\bibnamefont{Baltog}},
  \bibinfo{journal}{Journal of Luminescence} \textbf{\bibinfo{volume}{182}}.
%  \bibinfo{pages}{166 } (\bibinfo{year}{2017}).

\bibitem[{\citenamefont{Saitoh et~al.}(2000)\citenamefont{Saitoh, Komatsu, and
  Karasawa}}]{SAITOH2000}
\bibinfo{author}{\bibfnamefont{A.}~\bibnamefont{Saitoh}},
  \bibinfo{author}{\bibfnamefont{T.}~\bibnamefont{Komatsu}}, \bibnamefont{and}
  \bibinfo{author}{\bibfnamefont{T.}~\bibnamefont{Karasawa}},
  \bibinfo{journal}{Journal of Luminescence} \textbf{\bibinfo{volume}{87-89}}.
%  \bibinfo{pages}{633 } (\bibinfo{year}{2000}).

\bibitem[{\citenamefont{Karasawa et~al.}(1981)\citenamefont{Karasawa, Komatsu,
  Miyata, Iida, and Kaifu}}]{KARASAWA1981}
\bibinfo{author}{\bibfnamefont{T.}~\bibnamefont{Karasawa}},
  \bibinfo{author}{\bibfnamefont{T.}~\bibnamefont{Komatsu}},
  \bibinfo{author}{\bibfnamefont{K.}~\bibnamefont{Miyata}},
  \bibinfo{author}{\bibfnamefont{T.}~\bibnamefont{Iida}}, \bibnamefont{and}
  \bibinfo{author}{\bibfnamefont{Y.}~\bibnamefont{Kaifu}},
  \bibinfo{journal}{Physica B+C} \textbf{\bibinfo{volume}{105}},
  \bibinfo{pages}{88 } (\bibinfo{year}{1981}).

\bibitem[{\citenamefont{Tiwari et~al.}(2018)\citenamefont{Tiwari, Alibhai, and
  Fermin}}]{Tiwari2018}
\bibinfo{author}{\bibfnamefont{D.}~\bibnamefont{Tiwari}},
  \bibinfo{author}{\bibfnamefont{D.}~\bibnamefont{Alibhai}}, \bibnamefont{and}
  \bibinfo{author}{\bibfnamefont{D.~J.} \bibnamefont{Fermin}},
  \bibinfo{journal}{ACS Energy Lett.} \textbf{\bibinfo{volume}{3}},
  \bibinfo{pages}{1882} (\bibinfo{year}{2018}).
%  \urlprefix\url{https://doi.org/10.1021/acsenergylett.8b01182}.

\bibitem[{\citenamefont{Scholz et~al.}(2018)\citenamefont{Scholz, Oum, and
  Lenzer}}]{Scholz2018}
\bibinfo{author}{\bibfnamefont{M.}~\bibnamefont{Scholz}},
  \bibinfo{author}{\bibfnamefont{K.}~\bibnamefont{Oum}}, \bibnamefont{and}
  \bibinfo{author}{\bibfnamefont{T.}~\bibnamefont{Lenzer}},
  \bibinfo{journal}{Phys. Chem. Chem. Phys.} \textbf{\bibinfo{volume}{20}},
  \bibinfo{pages}{10677} (\bibinfo{year}{2018}).
%  \urlprefix\url{https://doi.org/10.1039/C7CP07729G}.

\bibitem[{Sup()}]{Supplementary}
\bibinfo{note}{See Supplemental Material at [URL will be inserted by publisher].}

\bibitem[{\citenamefont{Brandt et~al.}(2015)\citenamefont{Brandt, Kurchin,
  Hoye, Poindexter, Wilson, Sulekar, Lenahan, Yen, Stevanovi?, Nino
  et~al.}}]{Brandt2015}
\bibinfo{author}{\bibfnamefont{R.~E.} \bibnamefont{Brandt}},
  \bibinfo{author}{\bibfnamefont{R.~C.} \bibnamefont{Kurchin}},
  \bibinfo{author}{\bibfnamefont{R.~L.~Z.} \bibnamefont{Hoye}},
  \bibinfo{author}{\bibfnamefont{J.~R.} \bibnamefont{Poindexter}},
  \bibinfo{author}{\bibfnamefont{M.~W.~B.} \bibnamefont{Wilson}},
  \bibinfo{author}{\bibfnamefont{S.}~\bibnamefont{Sulekar}},
  \bibinfo{author}{\bibfnamefont{F.}~\bibnamefont{Lenahan}},
  \bibinfo{author}{\bibfnamefont{P.~X.~T.} \bibnamefont{Yen}},
  \bibinfo{author}{\bibfnamefont{V.}~\bibnamefont{Stevanovi}},
  \bibinfo{author}{\bibfnamefont{J.~C.} \bibnamefont{Nino}},
  \bibnamefont{et~al.}, \bibinfo{journal}{J. Phys. Chem. Lett.}
  \textbf{\bibinfo{volume}{6}}, \bibinfo{pages}{4297} (\bibinfo{year}{2015}).
%  \urlprefix\url{https://doi.org/10.1021/acs.jpclett.5b02022}.

\bibitem[{\citenamefont{Habe and Nakamura}(2021)}]{Habe2021}
  \bibinfo{author}{\bibfnamefont{T.}~\bibnamefont{Habe}} \bibnamefont{and}
  \bibinfo{author}{\bibfnamefont{K.}~\bibnamefont{Nakamura}},
  \bibinfo{journal}{Phys. Rev. B} \textbf{\bibinfo{volume}{103}},
  \bibinfo{pages}{115409} (\bibinfo{year}{2021}).
%  \urlprefix\url{https://link.aps.org/doi/10.1103/PhysRevB.103.115409}.

\bibitem[{\citenamefont{Jellison et~al.}(1999)\citenamefont{Jellison, Ramey,
  and Boatner}}]{Jellison1999}
\bibinfo{author}{\bibfnamefont{G.~E.} \bibnamefont{Jellison}},
  \bibinfo{author}{\bibfnamefont{J.~O.} \bibnamefont{Ramey}}, \bibnamefont{and}
  \bibinfo{author}{\bibfnamefont{L.~A.} \bibnamefont{Boatner}},
  \bibinfo{journal}{Phys. Rev. B} \textbf{\bibinfo{volume}{59}},
  \bibinfo{pages}{9718} (\bibinfo{year}{1999}).
%  \urlprefix\url{https://doi.org/10.1103/PhysRevB.59.9718}.

\bibitem[{\citenamefont{Podraza et~al.}(2013)\citenamefont{Podraza, Qiu,
  Hinojosa, Xu, Motyka, Phillpot, Baciak, Trolier-McKinstry, and
  Nino}}]{Podraza2013}
\bibinfo{author}{\bibfnamefont{N.~J.} \bibnamefont{Podraza}},
  \bibinfo{author}{\bibfnamefont{W.}~\bibnamefont{Qiu}},
  \bibinfo{author}{\bibfnamefont{B.~B.} \bibnamefont{Hinojosa}},
  \bibinfo{author}{\bibfnamefont{H.}~\bibnamefont{Xu}},
  \bibinfo{author}{\bibfnamefont{M.~A.} \bibnamefont{Motyka}},
  \bibinfo{author}{\bibfnamefont{S.~R.} \bibnamefont{Phillpot}},
  \bibinfo{author}{\bibfnamefont{J.~E.} \bibnamefont{Baciak}},
  \bibinfo{author}{\bibfnamefont{S.}~\bibnamefont{Trolier-McKinstry}},
  \bibnamefont{and} \bibinfo{author}{\bibfnamefont{J.~C.} \bibnamefont{Nino}},
  \bibinfo{journal}{J. Appl. Phys.} \textbf{\bibinfo{volume}{114}},
  \bibinfo{pages}{033110} (\bibinfo{year}{2013}).
%  \urlprefix\url{https://doi.org/10.1063/1.4813486}.

\bibitem[{\citenamefont{Cheng et~al.}(1991)\citenamefont{Cheng, Vidal, Zeiger,
  Dresselhaus, Dresselhaus, and Ippen}}]{Cheng1991}
\bibinfo{author}{\bibfnamefont{T.~K.} \bibnamefont{Cheng}},
  \bibinfo{author}{\bibfnamefont{J.}~\bibnamefont{Vidal}},
  \bibinfo{author}{\bibfnamefont{H.~J.} \bibnamefont{Zeiger}},
  \bibinfo{author}{\bibfnamefont{G.}~\bibnamefont{Dresselhaus}},
  \bibinfo{author}{\bibfnamefont{M.~S.} \bibnamefont{Dresselhaus}},
  \bibnamefont{and} \bibinfo{author}{\bibfnamefont{E.~P.} \bibnamefont{Ippen}},
  \bibinfo{journal}{Appl. Phys. Lett.} \textbf{\bibinfo{volume}{59}},
  \bibinfo{pages}{1923} (\bibinfo{year}{1991}).
%  \urlprefix\url{https://doi.org/10.1063/1.106187}.

\bibitem[{\citenamefont{Zeiger et~al.}(1992)\citenamefont{Zeiger, Vidal, Cheng,
  Ippen, Dresselhaus, and Dresselhaus}}]{Zeiger1992}
\bibinfo{author}{\bibfnamefont{H.~J.} \bibnamefont{Zeiger}},
  \bibinfo{author}{\bibfnamefont{J.}~\bibnamefont{Vidal}},
  \bibinfo{author}{\bibfnamefont{T.~K.} \bibnamefont{Cheng}},
  \bibinfo{author}{\bibfnamefont{E.~P.} \bibnamefont{Ippen}},
  \bibinfo{author}{\bibfnamefont{G.}~\bibnamefont{Dresselhaus}},
  \bibnamefont{and} \bibinfo{author}{\bibfnamefont{M.~S.}
  \bibnamefont{Dresselhaus}}, \bibinfo{journal}{Phys. Rev. B}
  \textbf{\bibinfo{volume}{45}}, \bibinfo{pages}{768} (\bibinfo{year}{1992}).
  %\urlprefix\url{https://doi.org/10.1103/PhysRevB.45.768}.

\bibitem[{\citenamefont{Ishioka et~al.}(2006)\citenamefont{Ishioka, Kitajima,
  and Misochko}}]{Ishioka2006}
\bibinfo{author}{\bibfnamefont{K.}~\bibnamefont{Ishioka}},
  \bibinfo{author}{\bibfnamefont{M.}~\bibnamefont{Kitajima}}, \bibnamefont{and}
  \bibinfo{author}{\bibfnamefont{O.~V.} \bibnamefont{Misochko}},
  \bibinfo{journal}{Journal of Applied Physics} \textbf{\bibinfo{volume}{100}}.
  \bibinfo{pages}{093501} (\bibinfo{year}{2006}).
  %\urlprefix\url{https://doi.org/10.1063/1.2363746}.

\bibitem[{\citenamefont{Ishioka and Misochko}(2010)}]{Ishioka2010}
\bibinfo{author}{\bibfnamefont{K.}~\bibnamefont{Ishioka}} \bibnamefont{and}
  \bibinfo{author}{\bibfnamefont{O.~V.} \bibnamefont{Misochko}},
  \emph{\bibinfo{title}{Coherent Lattice Oscillations in Solids and Their
  Optical Control}} (\bibinfo{publisher}{Springer Berlin Heidelberg},
  \bibinfo{address}{Berlin, Heidelberg}, \bibinfo{year}{2010}), pp.
  \bibinfo{pages}{23--46}, ISBN \bibinfo{isbn}{978-3-642-03825-9}.
  %\urlprefix\url{https://doi.org/10.1007/978-3-642-03825-9_2}.

\bibitem[{\citenamefont{Pagliara et~al.}(2011)\citenamefont{Pagliara,
  Galimberti, Mor, Montagnese, Ferrini, Grandi, Galinetto, and
  Parmigiani}}]{Pagliara2011}
\bibinfo{author}{\bibfnamefont{S.}~\bibnamefont{Pagliara}},
  \bibinfo{author}{\bibfnamefont{G.}~\bibnamefont{Galimberti}},
  \bibinfo{author}{\bibfnamefont{S.}~\bibnamefont{Mor}},
  \bibinfo{author}{\bibfnamefont{M.}~\bibnamefont{Montagnese}},
  \bibinfo{author}{\bibfnamefont{G.}~\bibnamefont{Ferrini}},
  \bibinfo{author}{\bibfnamefont{M.~S.} \bibnamefont{Grandi}},
  \bibinfo{author}{\bibfnamefont{P.}~\bibnamefont{Galinetto}},
  \bibnamefont{and}
  \bibinfo{author}{\bibfnamefont{F.}~\bibnamefont{Parmigiani}},
  \bibinfo{journal}{J. Am. Chem. Soc.} \textbf{\bibinfo{volume}{133}},
  \bibinfo{pages}{6318} (\bibinfo{year}{2011}).
%  \urlprefix\url{https://doi.org/10.1021/ja1110738}.

\bibitem[{\citenamefont{Chernikov et~al.}(2015)\citenamefont{Chernikov,
  Ruppert, Hill, Rigosi, and Heinz}}]{Chernikov2015}
\bibinfo{author}{\bibfnamefont{A.}~\bibnamefont{Chernikov}},
  \bibinfo{author}{\bibfnamefont{C.}~\bibnamefont{Ruppert}},
  \bibinfo{author}{\bibfnamefont{H.}~\bibnamefont{Hill}},
  \bibinfo{author}{\bibfnamefont{A.~F.} \bibnamefont{Rigosi}},
  \bibnamefont{and} \bibinfo{author}{\bibfnamefont{T.~F.} \bibnamefont{Heinz}},
  \bibinfo{journal}{Nat. Photonics} \textbf{\bibinfo{volume}{9}},
  \bibinfo{pages}{466} (\bibinfo{year}{2015}).
  %\urlprefix\url{https://doi.org/10.1038/nphoton.2015.104}.

\bibitem[{\citenamefont{Pogna et~al.}(2016)\citenamefont{Pogna, Marsili,
  De~Fazio, Dal~Conte, Manzoni, Sangalli, Yoon, Lombardo, Ferrari, Marini
  et~al.}}]{Pogna2016}
\bibinfo{author}{\bibfnamefont{E.~A.~A.} \bibnamefont{Pogna}},
  \bibinfo{author}{\bibfnamefont{M.}~\bibnamefont{Marsili}},
  \bibinfo{author}{\bibfnamefont{D.}~\bibnamefont{De~Fazio}},
  \bibinfo{author}{\bibfnamefont{S.}~\bibnamefont{Dal~Conte}},
  \bibinfo{author}{\bibfnamefont{C.}~\bibnamefont{Manzoni}},
  \bibinfo{author}{\bibfnamefont{D.}~\bibnamefont{Sangalli}},
  \bibinfo{author}{\bibfnamefont{D.}~\bibnamefont{Yoon}},
  \bibinfo{author}{\bibfnamefont{A.}~\bibnamefont{Lombardo}},
  \bibinfo{author}{\bibfnamefont{A.~C.} \bibnamefont{Ferrari}},
  \bibinfo{author}{\bibfnamefont{A.}~\bibnamefont{Marini}},
  \bibnamefont{et~al.}, \bibinfo{journal}{ACS Nano}
  \textbf{\bibinfo{volume}{10}}, \bibinfo{pages}{1182} (\bibinfo{year}{2016}).
  %\urlprefix\url{https://doi.org/10.1021/acsnano.5b06488}.

\bibitem[{\citenamefont{Mor et~al.}(2017)\citenamefont{Mor, Herzog,
  Gole\ifmmode~\check{z}\else \v{z}\fi{}, Werner, Eckstein, Katayama, Nohara,
  Takagi, Mizokawa, Monney et~al.}}]{Mor2017}
\bibinfo{author}{\bibfnamefont{S.}~\bibnamefont{Mor}},
  \bibinfo{author}{\bibfnamefont{M.}~\bibnamefont{Herzog}},
  \bibinfo{author}{\bibfnamefont{D.}~\bibnamefont{Gole\ifmmode~\check{z}\else
  \v{z}\fi{}}}, \bibinfo{author}{\bibfnamefont{P.}~\bibnamefont{Werner}},
  \bibinfo{author}{\bibfnamefont{M.}~\bibnamefont{Eckstein}},
  \bibinfo{author}{\bibfnamefont{N.}~\bibnamefont{Katayama}},
  \bibinfo{author}{\bibfnamefont{M.}~\bibnamefont{Nohara}},
  \bibinfo{author}{\bibfnamefont{H.}~\bibnamefont{Takagi}},
  \bibinfo{author}{\bibfnamefont{T.}~\bibnamefont{Mizokawa}},
  \bibinfo{author}{\bibfnamefont{C.}~\bibnamefont{Monney}},
  \bibnamefont{et~al.}, \bibinfo{journal}{Phys. Rev. Lett.}
  \textbf{\bibinfo{volume}{119}}, \bibinfo{pages}{086401}
  (\bibinfo{year}{2017}).
  %\urlprefix\url{https://doi.org/10.1103/PhysRevLett.119.086401}.

\bibitem[{\citenamefont{Cunningham et~al.}(2017)\citenamefont{Cunningham,
  Hanbicki, McCreary, and Jonker}}]{Cunningham2017}
\bibinfo{author}{\bibfnamefont{P.~D.} \bibnamefont{Cunningham}},
  \bibinfo{author}{\bibfnamefont{A.~T.} \bibnamefont{Hanbicki}},
  \bibinfo{author}{\bibfnamefont{K.~M.} \bibnamefont{McCreary}},
  \bibnamefont{and} \bibinfo{author}{\bibfnamefont{B.~T.}
  \bibnamefont{Jonker}}, \bibinfo{journal}{ACS Nano}
  \textbf{\bibinfo{volume}{11}}, \bibinfo{pages}{12601} (\bibinfo{year}{2017}).
%  \urlprefix\url{https://doi.org/10.1021/acsnano.7b06885}.

\bibitem[{\citenamefont{Jiang et~al.}(2018)\citenamefont{Jiang, Chen, Zheng,
  Xu, and Tang}}]{Jiang2018}
\bibinfo{author}{\bibfnamefont{T.}~\bibnamefont{Jiang}},
  \bibinfo{author}{\bibfnamefont{R.}~\bibnamefont{Chen}},
  \bibinfo{author}{\bibfnamefont{X.}~\bibnamefont{Zheng}},
  \bibinfo{author}{\bibfnamefont{Z.}~\bibnamefont{Xu}}, \bibnamefont{and}
  \bibinfo{author}{\bibfnamefont{Y.}~\bibnamefont{Tang}},
  \bibinfo{journal}{Opt. Express} \textbf{\bibinfo{volume}{26}},
  \bibinfo{pages}{859} (\bibinfo{year}{2018}).
%  \urlprefix\url{http://www.opticsexpress.org/abstract.cfm?URI=oe-26-2-859}.

\bibitem[{\citenamefont{Ullrich et~al.}(2014)\citenamefont{Ullrich, Xi, and
  Wang}}]{Ullrich2014}
\bibinfo{author}{\bibfnamefont{B.}~\bibnamefont{Ullrich}},
  \bibinfo{author}{\bibfnamefont{H.}~\bibnamefont{Xi}}, \bibnamefont{and}
  \bibinfo{author}{\bibfnamefont{J.~S.} \bibnamefont{Wang}},
  \bibinfo{journal}{J. Appl. Phys.} \textbf{\bibinfo{volume}{115}},
  \bibinfo{pages}{233503} (\bibinfo{year}{2014}).
%  \urlprefix\url{https://doi.org/10.1063/1.4883761}.

\bibitem[{\citenamefont{Franceschini et~al.}(2020)\citenamefont{Franceschini,
  Carletti, Pushkarev, Preda, Perri, Tognazzi, Ronchi, Ferrini, Pagliara, Banfi
  et~al.}}]{Franceschini2020}
\bibinfo{author}{\bibfnamefont{P.}~\bibnamefont{Franceschini}},
  \bibinfo{author}{\bibfnamefont{L.}~\bibnamefont{Carletti}},
  \bibinfo{author}{\bibfnamefont{A.~P.} \bibnamefont{Pushkarev}},
  \bibinfo{author}{\bibfnamefont{F.}~\bibnamefont{Preda}},
  \bibinfo{author}{\bibfnamefont{A.}~\bibnamefont{Perri}},
  \bibinfo{author}{\bibfnamefont{A.}~\bibnamefont{Tognazzi}},
  \bibinfo{author}{\bibfnamefont{A.}~\bibnamefont{Ronchi}},
  \bibinfo{author}{\bibfnamefont{G.}~\bibnamefont{Ferrini}},
  \bibinfo{author}{\bibfnamefont{S.}~\bibnamefont{Pagliara}},
  \bibinfo{author}{\bibfnamefont{F.}~\bibnamefont{Banfi}},
  \bibnamefont{et~al.}, \bibinfo{journal}{ACS Nano}
  \textbf{\bibinfo{volume}{14}}, \bibinfo{pages}{13602} (\bibinfo{year}{2020}).
%  \urlprefix\url{https://doi.org/10.1021/acsnano.0c05710}.

\bibitem[{\citenamefont{Jnawali et~al.}(2020)\citenamefont{Jnawali, Xiang,
  Linser, Shojaei, Wang, Qiu, Lian, Wong, Wu, Ye et~al.}}]{Jnawali2020}
\bibinfo{author}{\bibfnamefont{G.}~\bibnamefont{Jnawali}},
  \bibinfo{author}{\bibfnamefont{Y.}~\bibnamefont{Xiang}},
  \bibinfo{author}{\bibfnamefont{S.~M.} \bibnamefont{Linser}},
  \bibinfo{author}{\bibfnamefont{I.~A.} \bibnamefont{Shojaei}},
  \bibinfo{author}{\bibfnamefont{R.}~\bibnamefont{Wang}},
  \bibinfo{author}{\bibfnamefont{G.}~\bibnamefont{Qiu}},
  \bibinfo{author}{\bibfnamefont{C.}~\bibnamefont{Lian}},
  \bibinfo{author}{\bibfnamefont{B.~M.} \bibnamefont{Wong}},
  \bibinfo{author}{\bibfnamefont{W.}~\bibnamefont{Wu}},
  \bibinfo{author}{\bibfnamefont{P.~D.} \bibnamefont{Ye}},
  \bibnamefont{et~al.}, \bibinfo{journal}{Nat. Comm.}
  \textbf{\bibinfo{volume}{11}} (\bibinfo{year}{2020}).
%  \urlprefix\url{https://doi.org/10.1038/s41467-020-17766-5}.

\end{thebibliography}
\end{document}